\documentclass[11pt,superscriptaddress,aps,prd,preprint,showpacs]{revtex4}

\usepackage[dvips]{graphicx}
\usepackage{amsfonts}
\usepackage{slashed}
\usepackage[utf8x]{inputenc}
\usepackage{amsmath}
\usepackage{hyperref}

\begin{document}

\title{Generation of Carroll-Field-Jackiw term in the Functional Integral approach within Horava-Lifshitz $z=3$ CPT-violating QED}

\author{T. Mariz}
\affiliation{Instituto de F\'{\i}sica, Universidade Federal de Alagoas,\\ 57072-900, Macei\'o, Alagoas, Brazil}
\email{tmariz,rmartinez@fis.ufal.br}

\author{R. Martinez}
\affiliation{Instituto de F\'{\i}sica, Universidade Federal de Alagoas,\\ 57072-900, Macei\'o, Alagoas, Brazil}
\email{tmariz,rmartinez@fis.ufal.br}

\author{J. R. Nascimento}
\affiliation{Departamento de F\'{\i}sica, Universidade Federal da Para\'{\i}ba,\\
 Caixa Postal 5008, 58051-970, Jo\~ao Pessoa, Para\'{\i}ba, Brazil}
\email{jroberto,petrov@fisica.ufpb.br}

\author{A. Yu. Petrov}
\affiliation{Departamento de F\'{\i}sica, Universidade Federal da Para\'{\i}ba,\\
 Caixa Postal 5008, 58051-970, Jo\~ao Pessoa, Para\'{\i}ba, Brazil}
\email{jroberto,petrov@fisica.ufpb.br}

\begin{abstract}
In this paper, we apply the functional integral methodology to induce the Carroll-Field-Jackiw (CFJ) term in Horava-Lifshitz $z=3$ CPT-violating QED, where Lorentz and CPT breaking for fermion and photon sectors is introduced, and show that the CFJ term is finite but undetermined. 
\end{abstract}


\maketitle

\section{Introduction}	
	
Various Lorentz-breaking theories have been considered in many works in recent years. An important line of their study has been started by Horava, who proposed an original approach. It is based on the suggestion that, unlike the usual Lorentz-breaking theories constructed in a fashion described in \cite{ColKost}, whose action is a sum of some known Lorentz-invariant action and a small term proportional to a small constant Lorentz-breaking tensor, the Lorentz symmetry breaking is assumed to be large, so, in principle, the Lorentz symmetry can arise only in the low-energy limit of this theory. On the basis of this approach, the anisotropic theory of gravity \cite{horava} has been formulated. Another motivation for theories with strong Lorentz symmetry breaking is based on the anisotropic scaling originally proposed by Lifshitz within condensed matter physics \cite{Lifshitz}. The Lifshitz scaling is described by transformations $x\rightarrow bx, t\rightarrow b^zt$, where the integer $z$ is a so-called critical exponent. As a consequence, the resulting action involves two-time derivatives and $2z$ space ones. This allows to construct power-counting renormalizable gravity model since for $z=3$ the dimension of gravitational constant turns out to be zero as it is necessary for the renormalizability, while, as it is  known, the mass dimension of the gravitational constant in the usual Einstein gravity is $(-2)$ implying in its non-renormalizability known to be the main obstacle for attempts to construct a consistent quantum description of the gravity. Clearly, this also calls interest to develop the so-called Horava-Lifshitz (HL)-like extensions also for other field theory models. The most important examples within the context of this paper are HL-like theories with four-fermion interactions which have been studied for $z=3$ \cite{bumb2} and, in the Gross-Neveu case, for any $z$ \cite{HLGN} and the HL-like spinor QED that was considered for $z=2$  \cite{hlqed2} and for $z=3$ \cite{hlqed3}. 
	
One of the interesting problems within the study of HL theories is the possibility of consistent embedding these theories into the general Lorentz-breaking context allowing to establish relations between HL-like theories and ``usual" Lorentz-breaking theories formulated along the lines described in \cite{ColKost} and further papers. In principle, the possible aim of this study could consist in the development of a universalized description of Lorentz symmetry breaking involving HL-like theories and usual Lorentz-breaking theories as particular cases. One of the first steps in this way has been made in our previous paper \cite{hlcfj}, where we demonstrated how the Carroll-Field-Jackiw (CFJ) term can arise within the $z=3$ HL CPT-violating QED. In this paper, we generalize this study by applying the functional integral approach \cite{chung}.

The structure of the paper looks like follows. In section 2 we formulate our theory, introduce chiral transformations, and obtain corresponding variations of the action and the integral measure. In section 3 we perform the one-loop perturbative calculations. Section 4 is our Summary, where the results are discussed.

\section{Description of the theory}
	
	We start with the following CPT and Lorentz violating action   	
	 \begin{equation}\label{Lpsi}
	S = \int d^4x \left[ \bar\psi(i\slashed{D}_0+(i\slashed{D})^3-m^3-\slashed{b}_0\gamma_5+(\slashed{b}\slashed{D}\slashed{D})\gamma_5)\psi+e^2c\, b_\mu{^*F}^{\mu\nu}A_\nu \right],
	\end{equation} 
where $\slashed{D}_0=D_0\gamma^0$, $\slashed{D}=D_i\gamma^i$, $\slashed{b}_0=b_0\gamma^0$, $(\slashed{b}\slashed{D}\slashed{D})=(bDD)_{ijk}\gamma^i\gamma^j\gamma^k$, with
	 $D_\mu=\partial_\mu+ieA_\mu$, ${^*F}^{\mu\nu}=\frac{1}{2}\epsilon^{\mu\nu\lambda\rho}F_{\lambda\rho}$, and $c$ is an unknown constant introduced in a whole analogy with \cite{chung} to take into account the ambiguity in the definition of the conserved current $J^{\mu}_5$ whose explicit expression is not important, see the discussion in \cite{Bakas,Bakas1}. We note that, in principle, the exact form of the zero component of this current $J^0_5$, by dimensional reasons, should look like Eq. (2.9) from \cite{Bakas}, and of the spatial part of the current $J^i_5$ -- like Eq. (2.10) from \cite{Bakas}, plus, at most, some additive spinor-dependent terms involving $\gamma_5$ matrix as well as other terms in the same Eq. (2.10), and being of the second order in $b_i$, to be parity-odd. However, let us consider the continuity equation $\partial_{\mu}J^{\mu}_5=0$. If the current depends on $b_i$, it means that its $b$-independent part and its $b$-dependent part of the current should satisfy this equation separately since $b_i$ is a free parameter. Therefore, the $b$-independent part of the current is conserved itself, and we can deal only with it as with the whole current. Then, we can add to this current the additive term $e^2c\, {^*F}^{\mu\nu}A_\nu$, with the arbitrary $c$, so that we have now $\tilde{J}^{\mu}_5=J^{\mu}_5+e^2c\,{^*F}^{\mu\nu}A_\nu$ and, consequently, $\int dt d^3x\, \partial_{\mu}\tilde{J}^{\mu}_5=0$ as well. The action (\ref{Lpsi}) is chosen in order to break the CPT symmetry through introducing of $b_\mu$ axial vector so that the CFJ term $b_{\mu}{}^*F^{\mu\nu}A_{\nu}$ is consistent from the dimensional viewpoint.
	 In a certain sense, our theory can be treated as an extended Horava-Lifshitz analogue of the theory treated in \cite{Kapoor} where the axial vector was coupled to the fermionic field within the Lorentz-invariant framework. From another side, the methodology we follow here actually represents itself as an example of emergent dynamics used within various contexts, f.e. in \cite{Leblanc} it was used to generate effective dynamics for vector and antisymmetric tensor fields from their couplings to fermions within the Lorentz-invariant framework.
	 %{\it the weighted mass dimension of all its components is 1, just as occurs in the ``usual" Lorentz-breaking QED} {\bf(NA VERDADE $b_0$ TEM DIMENS\~AO $3$)}.
	 
	 Using the methodology applied in \cite{bumb2}, we introduce
	\begin{equation}
	(bDD)_{ijk}=\lambda_1b_iD_jD_k+\lambda_2b_jD_iD_k+\lambda_3b_kD_iD_j.
	\end{equation}
Since the action (\ref{Lpsi}) is completely expressed in terms of gauge covariant derivatives, it is naturally gauge invariant. It is more convenient to rewrite (\ref{Lpsi}) in the manner with explicit spatial and time derivatives:
	\begin{eqnarray}\label{Lpsi2}
	S &=& \int d^4x [ \bar\psi(i\slashed{\partial}_0+(i\slashed{\partial})^3-m^3-\slashed{b}_0\gamma_5+(\slashed{b}\slashed{\partial}\slashed{\partial})\gamma_5-e\slashed{A}_0+e(\slashed{\partial}\slashed{\partial}\slashed{A})+ie(\slashed{b}\slashed{\partial}\slashed{A})\gamma_5 \nonumber\\
	&&+ie^2(\slashed{\partial}\slashed{A}\slashed{A})-e^2(\slashed{b}\slashed{A}\slashed{A})\gamma_5-e^3\slashed{A}^3)\psi+e^2c\, b_\mu {^*F}^{\mu\nu}A_\mu],
	\end{eqnarray}
	with	\begin{equation}
	(b\partial\partial)_{ijk}=\lambda_1b_i\partial_j\partial_k+\lambda_2b_j\partial_i\partial_k+\lambda_3b_k\partial_i\partial_j,
	\end{equation}
	\begin{equation}
	(\partial\partial A)_{ijk} = (\partial_i\partial_jA_k)+(\partial_jA_k)\partial_i+(\partial_iA_k)\partial_j+A_k\partial_i\partial_j+(\partial_iA_j)\partial_k+A_j\partial_i\partial_k+A_i\partial_j\partial_k,
	\end{equation}
	\begin{eqnarray}
	(b\partial A)_{ijk} &=& \lambda_1b_i(\partial_jA_k)+\lambda_1b_iA_k\partial_j+\lambda_1b_iA_j\partial_k+ \lambda_2b_j(\partial_iA_k)+\lambda_2b_jA_k\partial_i+\lambda_2b_jA_i\partial_k \nonumber\\
	&&+\lambda_3b_k(\partial_iA_j)+\lambda_3b_kA_j\partial_i+\lambda_3b_kA_i\partial_j,
	\end{eqnarray}
	\begin{equation}
	(\partial AA)_{ijk} = (\partial_iA_j)A_k+A_j(\partial_iA_k)+A_jA_k\partial_i+A_i(\partial_jA_k)+A_iA_k\partial_j+A_iA_j\partial_k,
	\end{equation}
	and 
	\begin{equation}
	(bAA)_{ijk}=\lambda_1b_iA_jA_k+\lambda_2b_jA_iA_k+\lambda_3b_kA_iA_j.
	\end{equation}
Following the approach developed in \cite{bumb2}, we wrote this action for the general case when $\lambda_i$'s are not restricted. Further, we will fix them.
		
The generating functional, describing our theory, is given by
	\begin{equation} \label{act}
	Z = \int D\bar\psi D\psi\, e^{iS}.
	\end{equation}	
Now, let us follow the lines of \cite{chung} and perform the same change of variables as there, that is, chiral transformations:
\begin{eqnarray}
\psi(x) \rightarrow e^{i\alpha(x)\gamma_{5}}\psi(x),  \\
\bar{\psi}(x) \rightarrow \bar{\psi}(x) e^{i\alpha(x)\gamma_{5}}.
\end{eqnarray} 
Repeating all transformations performed in \cite{Bakas,Bakas1}, we can find that the integration measure of (\ref{act}) changes by  the same factor, as follows:
\begin{equation}\label{medida}
D\bar{\psi} D \psi \rightarrow D\bar{\psi} D \psi \:  \mathrm{exp} \left[ -i e^2 \int d^4x \frac{\alpha(x)}{8\pi^2} {}^*F^{\mu \nu} F_{\mu \nu} \right].
\end{equation} 
Actually, the theory we consider here is very similar to that one discussed in \cite{Bakas,Bakas1}, except of the $b_{\mu}$ and mass terms which, however, {cannot yield the contributions to the integral measure possessing the desired structure. Moreover, it has been argued in \cite{Bakas,Bakas1} that the anomaly is completely characterized by topology, and there is no differences between various contractions of spatial covariant derivatives, say $-\frac{i}{2}(D_i^2\slashed{D}+\slashed{D}D_i^2)$, or $-i D_i\slashed{D}D^i$, or other choices.  
%In the same paper, it has been argued that the explicit form of the conserved chiral current is not essential. 
In fact, using $\gamma^i\gamma^j\gamma^k = g^{ij}\gamma^k+g^{jk}\gamma^i-g^{ik}\gamma^j-i\epsilon^{0ijk}\gamma_0\gamma_5$, we can rewrite the spatial covariant derivatives of~(\ref{Lpsi}) as 
\begin{equation}
(i\slashed{D})^3=-iD^2_i\slashed{D}-i\slashed{D}D^2_i+iD_i\slashed{D}D^i-\epsilon^{0ijk}\gamma_0\gamma_5D_iD_jD_k,
\end{equation}
or better, taking into account that $[D_i,D_j]=ieF_{ij}$, as
\begin{equation}
(i\slashed{D})^3=-\frac{i}2D_i^2\slashed{D}-\frac{i}2\slashed{D}D_i^2-\frac{ie}2\partial_iF^{ij}\gamma_j-\frac{ie}2\epsilon^{0ijk}\gamma_0\gamma_5F_{ij}D_k.
\end{equation}
It is straightforward to see that the last two above terms do not contribute to the anomaly in~(\ref{medida}). So, we conclude that our contribution from variations of the measure is the same as in \cite{Bakas,Bakas1}.

Then, choosing $\alpha(x)=-x^\mu b_\mu$, similarly to \cite{chung}, and defining $\lambda_1=\lambda_2=\lambda_3=1$, for the sake of simplicity and in analogy with \cite{bumb2},  after disregarding the higher $b_\mu$ terms which is natural since $b_{\mu}$ is assumed to be small, we find that under the chiral transformations, our action goes to
	\begin{eqnarray}\label{Lpsi3}
	S &=& \int d^4x [ \bar\psi(i\slashed{\partial}_0+(i\slashed{\partial})^3-m^3-e\slashed{A}_0+e(\slashed{\partial}\slashed{\partial}\slashed{A})+ie^2(\slashed{\partial}\slashed{A}\slashed{A})-e^3\slashed{A}^3+
\nonumber\\&+&	2ix^\mu b_\mu\gamma_5m^3)\psi+e^2c\, b_\mu {^*F}^{\mu\nu}A_\mu],
	\end{eqnarray}
and, thus, our generating functional becomes
	\begin{eqnarray}\label{z}
	Z &=& \mathrm{exp} \left[-ie^2 \int \frac{d^4x}{4\pi^2} b_\mu {^*F}^{\mu \nu} A_\nu \right]  \mathrm{exp} \left[  ie^2c \int d^4x\, b_\mu {^*F}^{\mu\nu}A_\nu \right] \\ 
	&& \times \int D\bar{\psi} D \psi\, \mathrm{exp} \left[ i\int d^4x[\bar\psi(i\slashed{\partial}_0+(i\slashed{\partial})^3-m^3+2ix^\mu b_\mu\gamma_5m^3-e\slashed{A}_0+e(\slashed{\partial}\slashed{\partial}\slashed{A})+\right.\nonumber\\&&+\left.
	ie^2(\slashed{\partial}\slashed{A}\slashed{A})-e^3\slashed{A}^3)\psi]  \right] \nonumber. 
	\end{eqnarray}
This generating functional and the corresponding one-loop effective action will be studied in the next section.
	
	\section{Perturbative calculations}

Let us now calculate the CFJ term generated by the fermionic sector of (\ref{z}). By integrating out the spinor fields, we obtain the one-loop effective action
\begin{equation}\label{Seff}
\begin{split}
S_\mathrm{eff} =& -i\mathrm{Tr}\ln(\slashed{p}_0+\slashed{p}p_j^2-m^3+2ix^\mu b_\mu\gamma_5m^3 \\
&-e\slashed{A}_0-e\Delta_i(k,p)A^i+e^2\nabla^{ij}(k,p)A^iA^j-e^3\slashed{A}A_j^2),
\end{split}
\end{equation} where 
\begin{equation}
\Delta^i(k,p) = \slashed{k}\slashed{k}\gamma^i+\slashed{p}\slashed{k}\gamma^i+\slashed{k}\slashed{p}\gamma^i+\slashed{p}\slashed{p}\gamma^i+\slashed{k}\gamma^i\slashed{p}+\slashed{p}\gamma^i\slashed{p}+\gamma^i\slashed{p}\slashed{p},
\end{equation} and
\begin{equation}
\nabla^{ij}(k,p) = \slashed{k}\gamma^i\gamma^j+\slashed{k}\gamma^j\gamma^i+\slashed{p}\gamma^i\gamma^j+\gamma^j\slashed{k}\gamma^i+\gamma^i\slashed{p}\gamma^j+\gamma^i\gamma^j\slashed{p}.
\end{equation}
with $i\partial_jA^i(x)\to k_jA^i(k)$. Here, $\mathrm{Tr}$ stands for the trace over the Dirac matrices, together with the functional trace described by the integration in momentum and coordinate spaces.

In order to single out the quadratic terms in $A_\mu$ of the effective action,  we initially rewrite the expression (\ref{Seff}) in terms of expansion in number of vertices:
\begin{equation}
S_\mathrm{eff}=S_\mathrm{eff}^{(0)}+\sum_{n=1}^\infty S_\mathrm{eff}^{(n)},
\end{equation}
where $S_\mathrm{eff}^{\psi (0)}=-i\mathrm{Tr}\ln G^{-1}(p)$, the field independent part, and 
\begin{eqnarray}
S_\mathrm{eff}^{\psi(n)} &=& \frac{i}{n}\mathrm{Tr}[G(p)(e\slashed{A}_0+e\Delta_i(k,p)A^i-e^2\nabla^{ij}(k,p)A^iA^j+e^3\slashed{A}A_j^2)]^n,
\end{eqnarray}
with $G(p)=(\slashed{p}_0+\slashed{p}p_j^2-m^3-2\frac{\partial}{\partial p^\mu} b_\mu \gamma_5m^3)^{-1}$, where we have used the prescription $x^\mu=i\frac{\partial}{\partial p_\mu}$, which is standard for the derivative expansion methodology \cite{Pani}.

After evaluating the trace over the coordinate space, by using the commutation relation $A_\mu(x)G(p)=G(p-k)A_\mu(x)$ and the completeness relation of the momentum space, for the quadratic action $A_\mu$, we have the one-vertex contribution
\begin{equation}\label{Seff1}
S_\mathrm{eff}^{(1)}\big|_{A^2} = i\int d^4x \Pi_1^{ij}A_i(x)A_j(x),
\end{equation}
and the two-vertex contribution
\begin{equation}\label{Seff2}
S_\mathrm{eff}^{(2)}\big|_{A^2} = \frac i2 \int d^4x\, (\Pi_2^{00}A_0(x)A_0(x)+\Pi_3^{i0}A_i(x)A_0(x)+\Pi_4^{0j}A_0(x)A_j(x)+\Pi_5^{ij}A_i(x)A_j(x)).
\end{equation}
In the momentum space, we obtain
\begin{equation}\label{Seff1p}
S_\mathrm{eff}^{(1)}\big|_{A^2} = i\int \frac{d^4k}{(2\pi)^4} \Pi_1^{ij}A_i(k)A_j(-k),
\end{equation}
and
\begin{equation}\label{Seff2p}
S_\mathrm{eff}^{(2)}\big|_{A^2} = \frac i2 \int \frac{d^4k}{(2\pi)^4} (\Pi_2^{00}A_0(k)A_0(-k)+\Pi_3^{i0}A_i(k)A_0(-k)+\Pi_4^{0j}A_0(k)A_j(-k)+\Pi_5^{ij}A_i(k)A_j(-k)),
\end{equation}
with
\begin{subequations}\label{Pi}
	\begin{eqnarray}
	\label{Pi1ij}
	\Pi_1^{ij} &=& -e^2\int\frac{d^{4}p}{(2\pi)^4}\mathrm{tr}\,G(p)\nabla^{ij}(k,p),\\
	\label{Pi200}\Pi_2^{00} &=& e^2\int\frac{d^{4}p}{(2\pi)^4}\mathrm{tr}\,G(p)\gamma^0G(p-k)\gamma^0,\\
	\label{Pi3i0}\Pi_3^{i0} &=& e^2\int\frac{d^{4}p}{(2\pi)^4}\mathrm{tr}\,G(p)\Delta^i(k,p)G(p-k)\gamma^0, \\
	\label{Pi40j}\Pi_4^{0j} &=& e^2\int\frac{d^{4}p}{(2\pi)^4}\mathrm{tr}\,G(p)\gamma^0G(p-k)\Delta^j(-k,p-k), \\
	\label{Pi5ij}\Pi_5^{ij} &=& e^2\int\frac{d^{4}p}{(2\pi)^4}\mathrm{tr}\,G(p)\Delta^i(k,p)G(p-k)\Delta^j(-k,p-k).
	\end{eqnarray}
\end{subequations} 
The term $\Pi^{00}_2$ does not contribute to CFJ term. 

 Regarding the other terms of  (\ref{Pi}), if we consider the expansion $G(p)=S(p)+2m^3S(p)\gamma_5 b_\mu\frac{\partial}{\partial p_\mu} S(p)+\cdots$, where the dots are for higher $b_\mu$ terms, with $S(p)=(\slashed{p}_0+\slashed{p}p_j^2-m^3)^{-1}$, we have seven contributions to the CFJ term, namely, $\Pi_1^{ij}=\Pi_{1a}^{ij} +\cdots$, $\Pi_3^{i0}=\Pi_{3a}^{i0}+\Pi_{3b}^{i0}+\cdots$, $\Pi_4^{i0}=\Pi_{4a}^{0j}+\Pi_{4b}^{0j}+\cdots$ and $\Pi_5^{i0}=\Pi_{5a}^{ij}+\Pi_{5b}^{ij}+\cdots$, with
\begin{subequations}
	\begin{eqnarray}
		\label{Piij1}\Pi_{1a}^{ij} &=& -2m^3e^2\int\frac{d^{4}p}{(2\pi)^4}\mathrm{tr}\,S(p)\gamma_5 b_\mu \frac{\partial}{\partial p_\mu}S(p)\nabla^{ij}(k,p),\\
	\label{Pi3i0a}\Pi_{3a}^{i0} &=& 2m³e^2\int\frac{d^{4}p}{(2\pi)^4}\mathrm{tr}\,S(p)\gamma_5b_\mu \frac{\partial}{\partial p_\mu} S(p) \Delta^i(k,p)S(p-k)\gamma^0, \\
	\label{Pi3i0b}\Pi_{3b}^{i0} &=& 2m³e^2\int\frac{d^{4}p}{(2\pi)^4}\mathrm{tr}\,S(p) \Delta^i(k,p)S(p-k)\gamma_5b_\mu \frac{\partial}{\partial p_\mu} S(p-k)\gamma^0, \\
		\label{Pi40ja}\Pi_{4a}^{0j} &=& 2m^3e^2\int\frac{d^{4}p}{(2\pi)^4}\mathrm{tr}\,S(p)\gamma^0S(p-k)\gamma_5 b_\mu \frac{\partial}{\partial p_\mu} S(p-k)\Delta^j(-k,p-k), \\
		\label{Pi40jb}\Pi_{4b}^{0j} &=& 2m^3e^2\int\frac{d^{4}p}{(2\pi)^4}\mathrm{tr}\,S(p)\gamma_5b_\mu \frac{\partial}{\partial p_\mu} S(p)\gamma^0S(p-k)\Delta^j(-k,p-k), \\
	\label{Pi5ija}\Pi_{5a}^{ij} &=& 2m^3e^2\int\frac{d^{4}p}{(2\pi)^4}\mathrm{tr}\,S(p)\gamma_5b_\mu \frac{\partial}{\partial p_\mu}S(p)\Delta^i(k,p)S(p-k)\Delta^j(-k,p-k),  \\
	\label{Pi5ijb}\Pi_{5b}^{ij} &=& 2m^3e^2\int\frac{d^{4}p}{(2\pi)^4}\mathrm{tr}\,S(p)\Delta^i(k,p)S(p-k)\gamma_5 b_\mu \frac{\partial}{\partial p_\mu}S(p-k)\Delta^j(-k,p-k).
	\end{eqnarray}
\end{subequations}
Other terms, denoted by dots, are irrelevant for our purposes, since they involve higher orders in $b_{\mu}$.

We are now going to apply the momentum derivative to the right, where we must consider
 \begin{equation}
 b_\mu \frac{\partial}{\partial p_\mu}S(p)=-S(p)(\slashed{b}_0+b_k\xi^k(p)) S(p),
 \end{equation}
 with $\xi^k(p) = \gamma^k p_i^2+\slashed{p}\gamma^k\slashed{p}+p_i^2\gamma^k$,
\begin{equation}
b_\mu \frac{\partial}{\partial p_\mu}\nabla^{ij}(k,p)=b_k\gamma^{ijk},
\end{equation}
with $\gamma^{ijk}=\gamma^i\gamma^j\gamma^k+\gamma^i\gamma^k\gamma^j+\gamma^k\gamma^i\gamma^j$,
\begin{equation}
b_\mu \frac{\partial}{\partial p_\mu}\Delta^i(k,p)=b_k\tilde{\nabla}^{ik}(k,p),
\end{equation}
with $\tilde{\nabla}^{ik}(k,p)=\gamma^k\slashed{k}\gamma^i+\slashed{k}\gamma^k\gamma^i+\gamma^k\slashed{p}\gamma^i+\slashed{p}\gamma^k\gamma^i+\slashed{k}\gamma^i\gamma^k+\gamma^k\gamma^i\slashed{p}+\slashed{p}\gamma^i\gamma^k+\gamma^i\gamma^k\slashed{p}+\gamma^i\slashed{p}\gamma^k$,
\begin{equation}
b_\mu \frac{\partial}{\partial p_\mu}S(p-k)=-S(p-k)(\slashed{b}_0+b_k\xi^k(p-k)) S(p-k),
\end{equation}
and, finally, 
\begin{equation}
b_\mu \frac{\partial}{\partial p_\mu}\Delta^i(-k,p-k)=b_k\tilde{\nabla}^{ik}(p-k,-k).
\end{equation}

Then, we can write $\Pi_{1a}^{ij}=\Pi_{1a1}^{ij}+\Pi_{1a2}^{ij}$, where
\begin{subequations}
	\begin{eqnarray}
	\Pi_{1a1}^{ij} &=& 2m^3e^2\int\frac{d^{4}p}{(2\pi)^4}\mathrm{tr}\,S(p)\gamma_5 S(p)(\slashed{b}_0+b_k\xi^k(p)) S(p) \nabla^{ij}(k,p),\\
	\Pi_{1a2}^{ij} &=& -2m^3e^2\int\frac{d^{4}p}{(2\pi)^4}\mathrm{tr}\,S(p)\gamma_5 S(p)b_k\gamma^{ijk},
	\end{eqnarray}
\end{subequations}
$\Pi_{3a}^{i0}=\Pi_{3a1}^{i0}+\Pi_{3a2}^{i0}+\Pi_{3a3}^{i0}$ and $\Pi_{3b}^{i0}=\Pi_{3b1}^{i0}$, with
\begin{subequations}
	\begin{eqnarray}
	\Pi_{3a1}^{i0} &=& -2m^3e^2\int\frac{d^{4}p}{(2\pi)^4}\mathrm{tr}\,S(p)\gamma_5 S(p)(\slashed{b}_0+b_k\xi^k(p)) S(p) \Delta^i(k,p)S(p-k)\gamma^0,\\
	\Pi_{3a2}^{i0} &=& 2mA^3e^2\int\frac{d^{4}p}{(2\pi)^4}\mathrm{tr}\,S(p)\gamma_5S(p) b_k\tilde\nabla^{ik}(k,p) S(p-k)\gamma^0, \\
	\Pi_{3a3}^{i0} &=& -2mA^3e^2\int\frac{d^{4}p}{(2\pi)^4}\mathrm{tr}\,S(p)\gamma_5 S(p) \Delta^i(k,p)S(p-k)(\slashed{b}_0+b_k\xi^k(p-k)) S(p-k)\gamma^0,\\
 \Pi_{3b1}^{i0} &=& -2mA^3e^2\int\frac{d^{4}p}{(2\pi)^4}\mathrm{tr}\,S(p) \Delta^i(k,p)S(p-k)\gamma_5S(p-k)(\slashed{b}_0+b_k\xi^k(p-k))S(p-k)\gamma^0, \hspace{1cm}
	\end{eqnarray}
\end{subequations}
$\Pi_{4a}^{0j}=\Pi_{4a1}^{0j}+\Pi_{4a2}^{0j}$ and $\Pi_{4b}^{0j}=\Pi_{4b1}^{0j}+\Pi_{4b2}^{0j}+\Pi_{4b3}^{0j}$ where
\begin{subequations}
	\begin{eqnarray}
\Pi_{4a1}^{0j} &=& -2m^3e^2\int\frac{d^{4}p}{(2\pi)^4}\mathrm{tr}\,S(p)\gamma^0S(p-k)\gamma_5 S(p-k)(\slashed{b}_0+b_k\xi^k(p-k)) \nonumber\\
&&\times S(p-k)\Delta^j(-k,p-k),\\
	\Pi_{4a2}^{0j} &=& 2m^3e^2\int\frac{d^{4}p}{(2\pi)^4}\mathrm{tr}\,S(p)\gamma^0S(p-k)\gamma_5 S(p-k)b_k\tilde\nabla^{jk}(-k,p-k),\\
	\Pi_{4b1}^{0j} &=& -2m^3e^2\int\frac{d^{4}p}{(2\pi)^4}\mathrm{tr}\,S(p)\gamma_5 S(p)(\slashed{b}_0+b_k\xi^k(p)) S(p) \gamma^0S(p-k)\Delta^j(-k,p-k),\\
	\Pi_{4b2}^{0j} &=&-2m^3e^2\int\frac{d^{4}p}{(2\pi)^4}\mathrm{tr}\,S(p)\gamma_5 S(p)\gamma^0 S(p-k)(\slashed{b}_0+b_k\xi^k(p-k)) \nonumber\\
	&&\times S(p-k)\Delta^j(-k,p-k), \\
	\Pi_{4b3}^{0j} &=&2m^3e^2\int\frac{d^{4}p}{(2\pi)^4}\mathrm{tr}\,S(p)\gamma_5S(p)\gamma^0S(p-k)b_k\tilde\nabla^{jk}(-k,p-k),
	\end{eqnarray}
\end{subequations}
and, finally, $\Pi_{5a}^{ij}=\Pi_{5a1}^{ij}+\Pi_{5a2}^{ij}+\Pi_{5a3}^{ij}+\Pi_{5a4}^{ij}$ and $\Pi_{5b}^{0j}=\Pi_{5b1}^{0j}+\Pi_{5b2}^{0j}$ with
\begin{subequations}
	\begin{eqnarray}
	\Pi_{5a1}^{ij} &=& -2m^3e^2\int\frac{d^{4}p}{(2\pi)^4}\mathrm{tr}\,S(p)\gamma_5 S(p)(\slashed{b}_0+b_k\xi^k(p)) S(p) \Delta^i(k,p)S(p-k)\Delta^j(-k,p-k), \hspace{1cm} \\
	\Pi_{5a2}^{0ij} &=& 2m^3e^2\int\frac{d^{4}p}{(2\pi)^4}\mathrm{tr}\,S(p)\gamma_5S(p)b_k\tilde\nabla^{ik}(k,p)S(p-k)\Delta^j(-k,p-k),\\
	\Pi_{5a3}^{ij} &=& -2m^3e^2\int\frac{d^{4}p}{(2\pi)^4}\mathrm{tr}\,S(p)\gamma_5S(p)\Delta^i(k,p)S(p-k)\\ &&\nonumber \times (\slashed{b}_0+b_k\xi^k(p-k)) S(p-k)\Delta^j(-k,p-k),\\
	\Pi_{5a4}^{ij} &=& 2m^3e^2\int\frac{d^{4}p}{(2\pi)^4}\mathrm{tr}\,S(p)\gamma_5 S(p)\Delta^i(k,p)S(p-k) b_k\tilde\nabla^{jk}(-k,p-k), \\
	\Pi_{5b1}^{ij} &=& -2m^3e^2\int\frac{d^{4}p}{(2\pi)^4}\mathrm{tr}\,S(p)\Delta^i(k,p)S(p-k)\gamma_5 S(p-k) \nonumber\\
	&& \times(\slashed{b}_0+b_k\xi^k(p-k)) S(p-k)\Delta^j(-k,p-k), \\
	\Pi_{5b2}^{ij} &=& 2m^3e^2\int\frac{d^{4}p}{(2\pi)^4}\mathrm{tr}\,S(p)\Delta^i(k,p)S(p-k)\gamma_5 S(p-k) b_k\tilde\nabla^{jk}(-k,p-k).
	\end{eqnarray}
\end{subequations}

Next, we must consider the expansion
 \begin{equation} 
 S(p-k)=S(p)+S(p)\Xi(k,p) S(p)+...
\end{equation}
where $\Xi(k,p) = \slashed{k}_0+p_i^2\slashed{k}+\slashed{p}\slashed{k}\slashed{p}-\slashed{p}k_i^2+\slashed{k}p_i^2-\slashed{k}\slashed{p}\slashed{k}-k_i^2\slashed{p}+\slashed{k}^3$. Disregarding the higher $\Xi(k,p)$ terms and calculating the trace over the Dirac matrices, we arrive at
\begin{subequations}
	\begin{eqnarray}
	\Pi_{1}^{ij} &=& 8ie^2m^6\epsilon^{0kij}b_0k_k\int \frac{d^4p}{(2\pi)^4} \frac{1}{(p_0^2+p_i^6-m^6)^2}, \\
	\Pi_{3}^{i0} &=& 8ie^2m^6\epsilon^{jki0}b_jk_k\int \frac{d^4p}{(2\pi)^4} \frac{1}{(p_0^2+p_i^6-m^6)^2} \nonumber\\
	&&-80ie^2m^6\epsilon^{jki0}b_jk_k\int \frac{d^4p}{(2\pi)^4}\frac{p_i^6}{(p_0^2+p_i^6-m^6)^3},\\
	\Pi_{4}^{0j} &=&-48ie^2m^6\epsilon^{ik0j}b_ik_k\int \frac{d^4p}{(2\pi)^4}\frac{p_i^6}{(p_0^2+p_i^6-m^6)^3}, \\
	\Pi_{5}^{ij} &=&-112e^2m^6\epsilon^{0kij}b_0k_k\int \frac{d^4p}{(2\pi)^4}\frac{p_i^6}{(p_0^2+p_i^6-m^6)^3} \nonumber\\ 
	&& -48ie^2m^6\epsilon^{k0ij}b_kk_0\int \frac{d^4p}{(2\pi)^4}\frac{p_i^6}{(p_0^2+p_i^6-m^6)^3}.
	\end{eqnarray}
\end{subequations}
The above integrals are particular cases of a more general integral, given by
\begin{equation}\label{int}
I=\int^{\infty}_{-\infty}\frac{dp_0}{2\pi}\int\frac{d^3\vec{p}}{(2\pi)^3}\frac{(-1)^\beta\vec p^{2\beta}}{(p_0^2-\vec p^6-m^6)^\alpha}.
\end{equation}
Making a Wick rotation  ($p_0\rightarrow ip_0$) and performing the dimensional regularization, through the replacement $ d^3\vec{p}/(2\pi)^3\rightarrow \mu^{3-d}d^d\vec{p}/(2\pi)^d $, where $\mu$ is an arbitrary scale parameter, the integral (\ref{int}) becomes
\begin{equation}\label{int2}
I=i(-1)^{\alpha+\beta}\mu^{3-d}\int^{\infty}_{-\infty}\frac{dp_0}{2\pi}\int\frac{d^d \vec p}{(2\pi)^d}\frac{\vec p^{2\beta}}{(p_0^2+\vec p^6+m^6)^\alpha}.
\end{equation}
Then, performing the integral over $p_0$, we obtain
\begin{equation}\label{int3}
I=i\frac{(-1)^{\alpha+\beta}}{2\pi^{1/2}}\frac{\Gamma\left( \lambda-\frac{1}{2}\right)}{\Gamma(\lambda)}\mu^{3-d}\int\frac{d^d\vec p}{(2\pi)^d}\vec p^{2\beta}(\vec p^6+m^6)^{\frac{1}{2}-\lambda},
\end{equation}
and making the integral over $\vec p$, the Eq.~(\ref{int3}) results in   
\begin{equation}
I=i\frac{(-1)^{\alpha+\beta}}{6}m^{d+2\beta+3-6\lambda}\mu^{3-d}\frac{\pi^{\frac{d-1}{2}}}{(2\pi)^d}\frac{\Gamma\left( \frac{d}{6}-\frac{\beta}{3} \right) \Gamma \left( \lambda-\frac{1}{2}-\frac{d}{6}-\frac{\beta}{3}  \right) }{\Gamma(\lambda) \Gamma \left(\frac{d}{2}\right) }.
\end{equation} 
Therefore, we obtain
\begin{subequations}
 \begin{eqnarray}
 	\Pi_{1}^{ij} &=&
 	 - \frac{4}{3}m^{d-3}e^2 \frac{\pi^{\frac{d-1}{2}}}{(2\pi)^d}\frac{\Gamma\left(\frac{d}{6}\right) \Gamma\left(\frac{3}{2}-\frac{d}{6}  \right) }{\Gamma\left( \frac{d}{2} \right)}b_0k_k\epsilon^{0kij}, \\
 	\Pi_{3}^{i0} &=& \left(\frac{20d-24}{18} \right)m^{d-3}e^2\frac{\pi^{\frac{d-1}{2}}}{(2\pi)^d} \frac{\Gamma\left(\frac{d}{6}\right) \Gamma\left(\frac{3}{2}-\frac{d}{6}  \right)  }{\Gamma\left( \frac{d}{2} \right)} b_jk_k\epsilon^{jki0}, \\
 	\Pi_{4}^{0j} &=& \frac{4d}{6}m^{d-3}e^2\frac{\pi^{\frac{d-1}{2}}}{(2\pi)^d}\frac{\Gamma\left(\frac{d}{6}\right) \Gamma\left(\frac{3}{2}-\frac{d}{6}  \right) }{\Gamma\left( \frac{d}{2} \right)}b_ik_k\epsilon^{ik0j}, \\
 	\Pi_{5}^{ij} &=& \frac{4d}{6}m^{d-3}e^2\frac{\pi^{\frac{d-1}{2}}}{(2\pi)^d}\frac{\Gamma\left(\frac{d}{6}\right) \Gamma\left(\frac{3}{2}-\frac{d}{6}  \right) }{\Gamma\left( \frac{d}{2} \right)}\left(\frac{7}{3}b_0k_k\epsilon^{0kij}+b_kk_0\epsilon^{k0ij}\right).
 	 \end{eqnarray}
 \end{subequations}
 
With these results, by considering $S_\mathrm{eff}^{(1)}\big|_{A^2}+S_\mathrm{eff}^{(2)}\big|_{A^2}\rightarrow S_\mathrm{CFJ}$, we get
\begin{eqnarray}
S_\mathrm{CFJ}&=&\int \frac{d^4k}{(2\pi)^4}m^{d-3}e^2\frac{\pi^{\frac{d-1}{2}}}{(2\pi)^d}\frac{\Gamma\left(\frac{d}{6}\right) \Gamma\left(\frac{3}{2}-\frac{d}{6}  \right) }{\Gamma\left( \frac{d}{2} \right)}\nonumber \\ && \times \left[ \left( \frac{7d-12}{9} \right)b_0k_k\epsilon^{0kij} +\frac{20d-24}{36}b_jk_k\epsilon^{jki0} + \frac{4d}{12}b_ik_k\epsilon^{ik0j} + \frac{4d}{12}b_kk_0\epsilon^{k0ij} \right].
\end{eqnarray}
Thus, for $d=3$, we obtain
\begin{equation}\label{Scfj}
 S_\mathrm{CFJ}=\int\frac{d^4k}{(2\pi)^4}\frac{e^2}{4\pi^2} b_\mu A_\nu \partial_\alpha A_\beta \epsilon^{\mu \nu \alpha \beta}.
\end{equation}
This expression is non-ambiguous just as its analogue in the usual LV QED \cite{chung}. It can be written as
\begin{equation}\label{Scfj2}
S_\mathrm{CFJ}=\int \frac{d^4k}{(2\pi)^4}\frac{e^2}{4\pi^2} b_\mu {^*F}^{\mu\nu}(k) A_\nu(-k).
\end{equation}

Taking together all contributions to the generating functional (\ref{z}), we have the following effective action
 \begin{equation}
 S_\mathrm{A}=\int \frac{d^4k}{(2\pi)^4}\frac{e^2}{4\pi^2} b_\mu {^*F}^{\mu\nu}(k)(1-1+4\pi^2c) A_\nu(-k) +{\cal O}(k^2),
 \end{equation}
therefore,
 \begin{equation}
S_\mathrm{A}= e^2c \int \frac{d^4k}{(2\pi)^4} b_\mu {^*F}^{\mu\nu}(k)A_\nu(-k) +{\cal O}(k^2).
\end{equation}

We conclude our calculation with the statement that the CFJ term is finite and completely characterized by a totally undetermined constant $c$ arising from the classical action and describing the ambiguity in the definition of the conserved current corresponding to the chiral symmetry. The similarity of this conclusion to the situation that occurred  in the usual Lorentz-breaking spinor QED \cite{chung} is a highly nontrivial result since our theory is, due to the strong space-time anisotropy, completely different from the usual Lorentz-breaking QED.

\section{Summary}

In the Horava-Lifshitz CPT violating QED, we calculate the induced CFJ term, starting from an anisotropic action with Lorentz and CPT violating terms in the fermionic and gauge sector. Using the functional integral formalism and the dimensional regularization scheme, we showed that the CFJ term arising from the fermionic sector is finite but undetermined by the constant $c$. This situation is similar to that one occurring in the ``usual" LV QED \cite{chung}.  Let us discuss our results.

First of all, it means that the ambiguity of the CFJ term is not a consequence of a choice of a certain calculation scheme but a fundamental feature of the theory, which, in a whole analogy with \cite{JackAmb}, is related with some anomaly. It was argued in \cite{JackAmb} that in the usual LV QED, this is the Adler-Bell-Jackiw (ABJ) anomaly. In the $z=3$ we considered here, this is the triangle anomaly discussed in great detail in \cite{Bakas,Bakas1}. Therefore, we conclude that the correspondence between the ambiguity and the anomaly holds in $z=3$ HL-like theories as well. It is natural to expect that the same situation is replayed for any odd $z$, since at any even $z$, the triangle graph yields zero result \cite{MNP}, and the corresponding anomaly does not emerge.

Second, effectively we have demonstrated that HL-like theories and usual LV theories are nothing more as limits of a some more generic theory. Apparently, it is some consequence of the concept of the renormalization group flow, which in \cite{horava} has been used to relate Lorentz-invariant Einstein-Hilbert gravity with HL gravity, and it is natural to conclude that it can be used to relate also usual LV theories with HL theories.

Third, using argumentation given in \cite{Bakas,Bakas1} within the HL gravity context, it is natural to expect that the ambiguous CS-like $z=3$ term can be generated in the HL gravity as a quantum correction as well (probably, the same is valid for other odd $z$'s). Considering the CS term  in the HL gravity, which has been treated in \cite{horava} as its possible ingredient, could be the natural continuation of this study.

{\bf Acknowledgements.} R. M. is thankful to CAPES and FAPEAL for the financial support. The work by A. Yu. P. has been partially supported by the CNPq project No. 301562/2019-9.

\end{document}